\documentclass[
superscriptaddress,
amsmath,amssymb,
aps,
prl,
twocolumn,
floatfix]{revtex4}
\usepackage{graphicx}
\usepackage{amssymb,amsmath,bbm,braket,indentfirst,bm}
\usepackage{dcolumn}
\usepackage{color}
\usepackage{bigstrut}

\newcommand{\abs}[1]{\ensuremath \left|#1\right|}

\renewcommand\({\ensuremath \left(}
\renewcommand\){\ensuremath \right)}
\renewcommand\[{\ensuremath \left[}
\renewcommand\]{\ensuremath \right]}

\begin{document}

\title{Materials design from non-equilibrium steady states: driven
  graphene\\ as a tunable semiconductor with topological properties}

\author{Thomas~Iadecola}
\affiliation{Physics Department, Boston University, Boston, Massachusetts 02215, USA}

\author{David~Campbell}
\affiliation{Physics Department, Boston University, Boston, Massachusetts 02215, USA}

\author{Claudio~Chamon}
\affiliation{Physics Department, Boston University, Boston, Massachusetts 02215, USA}

\author{Chang-Yu~Hou}
\affiliation{Department of Physics and Astronomy, University of California at Riverside, Riverside, California 92521,USA}
\affiliation{Department of Physics, California Institute of Technology, Pasadena, California 91125, USA}

\author{Roman~Jackiw}
\affiliation{Department of Physics, Massachusetts Institute of Technology, Cambridge, Massachusetts 02139, USA}

\author{So-Young~Pi}
\affiliation{Physics Department, Boston University, Boston, Massachusetts 02215, USA}

\author{Silvia~Viola~Kusminskiy}
\affiliation{Dahlem Center for Complex Quantum Systems and Fachbereich Physik, Freie Universit\"at Berlin, 14195 Berlin, Germany}

\date{\today}

\begin{abstract}
  Controlling the properties of materials by driving them out of
  equilibrium is an exciting prospect that has only recently begun to
  be explored.  In this paper we give a striking theoretical example
  of such materials design: a tunable gap in monolayer
  graphene is generated by exciting a particular optical phonon. We
  show that the system reaches a steady state whose transport
  properties are the same as if the system had a static electronic
  gap, controllable by the driving amplitude. Moreover, the steady
  state displays topological phenomena: there are chiral edge
  currents, which circulate a fractional charge $e/2$ per rotation
  cycle, with frequency set by the optical phonon frequency.
\end{abstract}

\maketitle

Non-equilibrium quantum systems constitute a natural frontier in
physics that is only beginning to be probed by theory and experiment.
Non-equilibrium methods can be used to study \cite{averittreview} and control \cite{stockmannature,stockmanprl} the properties of
condensed matter systems.  Particularly exciting is the
possibility of engineering the properties of novel materials, like graphene, by driving them out of equilibrium, paving the way for applications to devices.
Graphene's gaplessness poses a critical challenge to such
applications, as the development of graphene-based semiconductors is
predicated on the ability to induce a gap.

In this Letter we present a theoretical study in graphene of this
non-equilibrium approach to materials design.  We demonstrate the
possibility of inducing a gap in monolayer graphene by the excitation of optical
phonon modes.  The gap is controlled by a
time-dependent Kekul\'e-pattern bond density wave, which appears in
the effective field theory as a complex-valued order parameter
$\Delta$ that rotates with the frequency $\Omega$ of the driven
phonon mode.  The time-dependence in this order parameter is completely removable by an
axial (valley) gauge transformation, which
can be viewed as a kind of ``boost" to a co-moving ``reference frame."  The gauge transformation has no effect on the coupling of the system to a heat bath, thereby guaranteeing thermal equilibration in the new frame, and leaves the fermion currents invariant.  This implies that the electric response of the system is equivalent to that
of one with a static gap; all non-equilibrium aspects of
the problem are removed and the system can be studied as if it were at
equilibrium.

The topological consequences of the Kekul\'e gap have
been studied in the static case, revealing that fractionally charged
states can emerge that are bound to vortices in the order parameter
$\Delta$ \cite{changyu2007}.  In the driven case, we show that
further topological phenomena arise: the system supports
chiral edge currents of magnitude $J_{\rm edge}=e\,\Omega/4\pi$, while the current in the bulk
vanishes.  These results suggest the possibility that
driven graphene could be used as a tunable semiconductor with
nontrivial topological properties.



Let us consider spinless electrons hopping on a honeycomb lattice
$\Lambda$ according to the {\it time-dependent} tight-binding
Hamiltonian
\begin{align}\label{tightbinding}
H=-\sum_{\bm r\in\Lambda_A}\sum_{j=1}^3\[t+\delta t_{\bm r,j}(\tau)\] \; a_{\bm r}^\dagger b_{\bm r+\bm  s_j}\ +\ \text{h.c.}\ ,
\end{align}
where $\tau$ is time and $a_{\bm r}^\dagger$ and $b_{\bm r+\bm s_j}^\dagger$ are
fermionic creation operators at sites $\bm r\in \Lambda_A$ and $\bm
r+\bm s_j\in\Lambda_B$, with $\Lambda_A$ and $\Lambda_B$ the two
triangular sublattices forming the hexagonal lattice $\Lambda$. The vectors $\bm s_j$ ($j=1,2,3$) connect a site $\bm r\in
\Lambda_A$ to its three nearest neighbors at $\bm r+\bm
s_j\in\Lambda_B$ located a distance $|\bm s_j|=d$ away. The uniform
hopping amplitudes $t$ are modulated by time- and site-dependent
perturbations $\delta t_{\bm r,j}(\tau)$. In the absence of such
perturbations ($\delta t_{\bm r,j}(\tau)=0$), the
Hamiltonian \eqref{tightbinding} can be diagonalized in momentum
space, and the single particle spectrum has two Dirac points at $\bm
k=\bm K_{\pm}=\pm\frac{4\pi}{3\sqrt{3}d}\(1,0\)$.

We shall now consider the perturbations
$\delta t_{\bm r,j}(\tau)$ that result from the excitation of the
highest-energy optical phonon modes at wavevectors $\bm K_\pm$ with
frequency $\Omega$. The atomic displacements from the
lattice sites $\bm r_{A,B}\in\Lambda_{A,B}$ are
\begin{align}\label{vectormodes}
\bm u^{A,B}_{\bm K_{\pm}}(\bm r_{A,B},\tau)&=c_\pm\ e^{i\bm r_{A,B}\cdot\bm K_{\pm}}e^{-i\Omega \tau}\ \bm{u}^{A,B}_{\pm}+\text{c.c.}
\end{align} 
The coefficients $c_\pm$ are the amplitudes of the excited waves. The
normal mode vectors $\bm u^{A,B}_{\pm}$ for the highest-energy optical
modes with frequency $\Omega$ at wavevectors $\bm K_\pm$ can be determined from a classical
analysis of the lattice displacements \cite{mahan,ando} and are given
by
\begin{align}\label{modevectors}
\bm u^A_{\pm}=\frac{1}{2}\begin{pmatrix}1\\\mp i\end{pmatrix}\indent\text{and}\indent\bm u^B_\pm=\frac{1}{2}\begin{pmatrix}1\\\pm i\end{pmatrix}\ .
\end{align}
To determine the form of the hopping modulations $\delta t_{\bm
  r,j}(\tau)$ resulting from the phonons, we consider the
changes in bond lengths due to the atomic displacements
\eqref{vectormodes} when either the mode at $\bm K_+$ or $\bm K_-$ is excited. For small displacements, the change in
the length $d_{\bm r,j}(\tau)$ of the bond connecting site $\bm r$ and
$\bm r+\bm s_j$ is~\cite{claudiosolitons}
\begin{align}
\frac{\delta d^\pm_{\bm r,j}(\tau)}{d}&\approx -\frac{\bm  s_j}{d}\cdot \[\frac{\bm u_{\bm K_\pm}^A(\bm r,\tau)}{d}-\frac{\bm u_{\bm K_\pm}^B(\bm r+\bm  s_j,\tau)}{d}\]
\label{bondlength1}
\end{align}
Substituting \eqref{vectormodes} and \eqref{modevectors} into
\eqref{bondlength1} and using $e^{i K_{\pm}\cdot \bm  s_j}=e^{\pm i\frac{2\pi}{3}(j-1)} $, one obtains
\begin{align}\label{bondlength2}
\frac{\delta d^\pm_{\bm r,j}(\tau)}{d}&=\pm i\,\frac{c_\pm^*}{d}\,\bm e^{i\bm K_\pm\cdot\bm  s_j}e^{\pm i\bm G\cdot \bm r}e^{\pm i\Omega \tau}+\text{c.c.}\ ,
\end{align}
where the vector $\bm G=\bm K_+-\bm K_-=2\bm K_+$ connects the two Dirac points.  The modulation in the hopping amplitude is related to the change in bond length through $\delta t_{\bm r,j}(\tau)/t=\alpha\,\delta
d_{\bm r, j}^{\pm}(\tau)/d$, where $\alpha\approx3.7$ is the dimensionless electron-phonon
coupling \cite{claudiosolitons}. The resulting $\delta t_{\bm r,j}(\tau)$ can be written as
\begin{align}\label{hoppings}
\delta t_{\bm r,j}(\tau)=\frac{1}{3}\Delta(\tau)\ e^{i\bm K_+\cdot \bm s_j}e^{i\bm G\cdot \bm r}+\text{c.c.}\ ,
\end{align}
where 
\begin{align}\label{Delta}
\Delta(\tau)=\begin{cases}
i\,3\alpha t \,\frac{c_+^*}{d}\;e^{+i\Omega \tau} & \text{for the }\bm{K}_+\text{ mode}\\
i\,3\alpha t \,\frac{c_-}{d}\;e^{-i\Omega \tau} &\text{for the }\bm{K}_-\text{ mode.}
\end{cases}
\end{align}
The hopping modulations \eqref{hoppings} have the form of a Kekul\'e distortion with an order parameter
$\Delta(\tau)$~\cite{changyu2007} that is
time-dependent. Therefore, exciting either the $\bm K_+$ or the $\bm
K_-$ mode independently yields a Kekul\'e order parameter that rotates
in time with frequency $\Omega$ in opposite directions for the two modes.

Without loss of generality, we henceforth consider the case where the
$\bm K_+$ mode is excited, and write $\Delta(\tau) =
\abs\Delta\;e^{i\phi(\tau)}$, where $\phi(\tau)=\Omega\tau+\varphi$. All the results for the $\bm K_-$ mode are obtained from those below by
taking $\Omega\to-\Omega$.



We study the consequences of this rotating order parameter in
the context of the effective Dirac field theory of the system, which
is valid in the limit where the fermions have relativistic
(hyperbolic) dispersion.  In order to ensure the validity of this
approximation we require $\abs\Delta/t \ll 1$ and
$\Omega/t\ll 1$, where the uniform hopping amplitude $t$ sets the
kinetic energy scale of the problem.  In this regime the Hamiltonian
\eqref{tightbinding} corresponds, to first order in a gradient
expansion, to the Dirac Lagrangian density \cite{changyu2007,chiralgauge}
\begin{align}\label{tdeplagrangian}
\mathcal L &= \bar\Psi\[\gamma^\mu(i\partial_\mu +\gamma_5A_{5\, \mu})- \abs\Delta\, e^{-i\gamma_5\phi(\tau)}\]\Psi\ ,
\end{align}
\noindent with $\mu=0,1,2,$ $\bar\Psi=\Psi^\dagger\gamma^0$ and $4\times4$ Dirac matrices
\begin{align*}
\gamma^0\equiv\begin{pmatrix}
0&\mathbbm 1\\\mathbbm 1&0
\end{pmatrix}\ ,\indent \gamma^i&\equiv\begin{pmatrix} 0&-\sigma_i\\\sigma_i&0 \end{pmatrix}\ ,\nonumber\\
 \indent\gamma_5\equiv i\gamma^0\gamma^1\gamma^2\gamma^3&=\begin{pmatrix}
\mathbbm 1&0\\0&-\mathbbm 1
\end{pmatrix}\ ,
\end{align*}
where $\mathbbm1$ is the $2\times 2$ unit matrix and $\sigma_i$ are
the three Pauli matrices.  The Dirac spinor $\Psi^\dagger_{\bm
  p}=(b_{\bm p,+}^\dagger\ a_{\bm p,+}^\dagger\ a_{\bm p,-}^\dagger\
b_{\bm p,-}^\dagger)$ collects the creation operators $a_{\bm
  p,\pm}^\dagger$ and $b_{\bm p,\pm}^\dagger$ for the $\pm$ species on
sublattices A and B, respectively. The axial gauge field $A_{5\,
  \mu}$, examined in a different context in Ref.~\onlinecite{chiralgauge}, plays an
important role in the discussion of the asymptotic steady state of the
driven system. The spatial components $A_{5\, i}$ correspond
physically to acoustic phonons and strain in the graphene lattice. If
the lattice is strained uniaxially, the hopping amplitudes change, and
the Dirac points shift away from $\bm K_\pm$. In this case, the
$A_{5\, i}$ acquire a non-zero average value. In addition,
acoustic phonons, either in-plane or out-of-plane, dynamically stretch
the bonds, leading to fluctuations of $A_{5\, i}$ around the average. These acoustic phonons
provide a thermal bath and their coupling to the electronic degrees of
freedom provides a system-bath interaction, which enables the system to reach an out-of-equilibrium steady state. 

We now observe that the time-dependent mass term in the Lagrangian
\eqref{tdeplagrangian} can be made constant by the axial
(valley) gauge transformation
\begin{align}\label{gaugetransformation}
\tilde{\Psi} &= e^{-i\gamma_5\frac{\Omega}{2}\tau}\,\Psi\ ,\hspace{.25cm} \tilde A_{5\, 0}= A_{5\, 0}-\frac{\Omega}{2}\ ,\hspace{.25cm} \tilde A_{5\, i}=A_{5\, i}\ ,
\end{align}

\noindent where $i=1,2$.  The transformed Lagrangian is found to be
\begin{align}\label{staticlagrangian}
\tilde{\mathcal L} = \bar{\tilde\Psi}\[\gamma^\mu(i\partial_\mu+\gamma_5\tilde A_{5\, \mu})-\abs\Delta e^{-i\gamma_5\varphi}\]\tilde\Psi\ ,
\end{align}
where we used $\{\gamma_5,\gamma^\mu\}=0$. This transformation maps the problem to a frame of
reference which is ``co-moving" with the Kekul\'e mass, so that the
Lagrangian is no longer explicitly dependent on time.

The vector current operator $j^\mu = {\bar\Psi\gamma^\mu\Psi}$, which
is associated with the electric response of the system, and the axial
current operator $j_5^\mu={\bar\Psi\gamma^\mu\gamma_5\Psi}$ are
invariant under \eqref{gaugetransformation}. Furthermore, the
spatial components $A_{5\, i}$ of the axial gauge field are also invariant
under \eqref{gaugetransformation}. Since we
have taken the fluctuations in $A_{5\, i}$ to act as a heat bath, we conclude that
this transformation leaves the bath invariant. Moreover, it also leaves the
system-bath coupling $ A_{5\, i}\,j_5^i$ invariant. Therefore the transformation \eqref{gaugetransformation} removes {\it all} time-dependences---those of the system, the bath, and the
system-bath interactions. The
remarkable consequence is that the non-equilibrium steady state of the
time-dependent system corresponds to a thermal equilibrium state in the co-moving
frame!

Consequently, the Hamiltonian $\mathcal H$ corresponding to the transformed Lagrangian \eqref{staticlagrangian}
can be analyzed in the time-independent
Schr\"odinger picture at thermal equilibrium.  $\mathcal H$
takes a particularly simple form in the absence of strain, in which
case $A_{5\, \mu}=0$, {\it i.e.} $\tilde{A}_{5\, 0}=-\Omega/2$ and $\tilde{A}_{5\, i}=0$:
 \begin{align}\label{statichamiltonian}
 \mathcal H &= \begin{pmatrix}
\bm\sigma\cdot \bm p+\frac{\Omega}{2}\;\mathbbm{1}&\abs\Delta e^{i\varphi}\;\mathbbm{1}\\\abs\Delta e^{-i\varphi}\;\mathbbm{1}&-\bm \sigma\cdot\bm p-\frac{\Omega}{2}\;\mathbbm{1}
\end{pmatrix}\ ,
 \end{align}
 where $\bm \sigma$ is the 2D vector of Pauli matrices and $\bm
 p = -i\bm\nabla$.  The eigenvalue problem $\mathcal H\psi =E\psi$ has been solved in~\cite{majorana} in the context of the superconducting proximity effect in topological insulators~\cite{fukane}; the four energy eigenvalues of the Hamiltonian
 \eqref{statichamiltonian} are given by
\begin{align}\label{eq:spectrum}
E_{\pm,\mp}=\pm\sqrt{(p\mp\Omega/2)^2+\abs\Delta^2}\; .
\end{align}  
Evidently the gauge transformation \eqref{gaugetransformation} maps
the time-dependent problem of Eq.\ \eqref{tdeplagrangian} to a time-independent problem with
an energy gap $2\abs\Delta$. 

It is important to observe that, because the vector current operator
$j^\mu$ is invariant under \eqref{gaugetransformation}, {\it all}
observables associated with $j^\mu$ can be calculated from the
static Lagrangian \eqref{staticlagrangian} without dealing with the original time-dependent mass. In particular, the
conductivity tensor $\sigma_{ij}$ obtained from the Kubo formula
written in terms of the current operator $j^\mu$ can be computed from \eqref{staticlagrangian}. Consequently, the
driven graphene system effectively behaves as a semiconductor with
a gap $2\abs\Delta$ tunable by the amplitude of the optical phonon
mode.


We shall next demonstrate that the rotating Kekul\'e mass in the
Lagrangian \eqref{tdeplagrangian} gives rise to topological phenomena
beyond those that have been found in the static case. To do this, we
follow \cite{2ddirac} in studying a variant of \eqref{tdeplagrangian}:
\begin{align}\label{currentslagrangian}
\mathcal L &= \bar\Psi\[\gamma^\mu(i\partial_\mu +\gamma_5A_{5\, \mu})- \abs{\Delta} \, e^{-i\gamma_5\phi}-\gamma^3\mu\]\Psi\ ,
\end{align}
where the scalar field $\mu=\mu(\bm x)$ corresponds to a staggered chemical
potential that establishes an energy imbalance between the sites of
$\Lambda_A$ and $\Lambda_B$. The Kekul\'e field
$\Delta=\abs{\Delta(\bm x)}e^{i\phi(\bm x,\tau)}$, where $\phi(\bm
x,\tau)=\Omega\tau+\varphi(\bm x)$, now carries an explicit spatial
dependence. The fields $\mu$ and $\Delta$ correspond to independent masses in the Lagrangian \eqref{currentslagrangian}, {\it i.e.} the total effective mass of the charge carriers
is $\sqrt{\mu^2+|\Delta|^2}$. The vector current density in the
presence of (space- and time-dependent) masses $\mu$ and
$\Delta$ is given by~\cite{2ddirac}
\begin{align}
\label{currentformula}
\langle j^\mu \rangle
&= 
e\frac{i}{2\pi}\epsilon^{\mu\alpha\beta}
\left\{
\partial_\alpha\chi^*\partial_\beta\chi\,
-
i\,\partial_\alpha\left[(1-2|\chi|^2) A_{5\, \beta}\right]
\right\},
\end{align}
where $e$ is the electron charge, $\epsilon^{\mu\alpha\beta}$ is the
Levi-Civita symbol, and the complex-valued auxiliary field $\chi\equiv\sin(\theta/2)\; e^{i\phi}$, where
\begin{align}\label{chidefinition}
\cos\theta&=\frac{\mu}{\sqrt{\mu^2+|\Delta|^2}},\quad
\sin\theta\;e^{i\phi}=\frac{\Delta}{\sqrt{\mu^2+|\Delta|^2}}\;,
\end{align}
with $0\le \theta< \pi$ and $0\le \phi< 2\pi$. Equations \eqref{currentformula} and
\eqref{chidefinition} form the basis of our discussion of the
topological currents resulting from the time-dependence of the Kekul\'e mass
term in \eqref{currentslagrangian}.  We use $\mu$ to define an edge, setting
$\mu\to 0$ in the bulk and using the limit $\abs\mu\to\infty$ to define an insulating region
outside the sample \footnote{In the limit $|\mu|\to\infty$, propagation into one sublattice costs infinite energy, while propagation into the other is blocked by the Pauli principle.}.

The current density in \eqref{currentformula} is gauge-invariant, so one can compute it in the reference frame where $\phi$ has a time dependence or in the co-moving frame where $\phi$ (and $\chi$) are time-independent. It follows that the
averaged charge and current densities are
\begin{subequations}
\begin{align}\label{chargedensity}
&\langle \,\rho\, \rangle =e
\frac{i}{2\pi}\,\epsilon^{0ij}\,
\partial_i\chi^*\partial_j\chi
= \langle \,\rho\, \rangle_{\rm static}
\\
\label{currentdensity}
&\langle\, \bm j\, \rangle 
= 
-e\frac{\Omega}{2\pi}\,{\hat{\bm z}}\times \bm\nabla |\chi(\bm x)|^2 \; ,
\end{align}
\end{subequations}
where $\hat{\bm z}$ is the unit vector perpendicular to the plane of the
sample.

Several observations are in order. First, the charge
density in the case of the time-dependent Kekul\'e mass is identical
to that in the static case. Second, the current density is non-vanishing and proportional to
the rotation frequency $\Omega$. Notice that the rotating mass breaks
time-reversal symmetry, and therefore it is possible to have a
non-vanishing current. Third, if $|\Delta|$ does not vary spatially, the current vanishes; this is the
case in the bulk of a uniform graphene sample, where we
take $|\Delta|$ to be constant. Fourth, there are necessarily edge
currents, which we shall now discuss in detail.

\begin{figure}[htbp]
\begin{center}
  \includegraphics[angle=90,width=.38\textwidth]{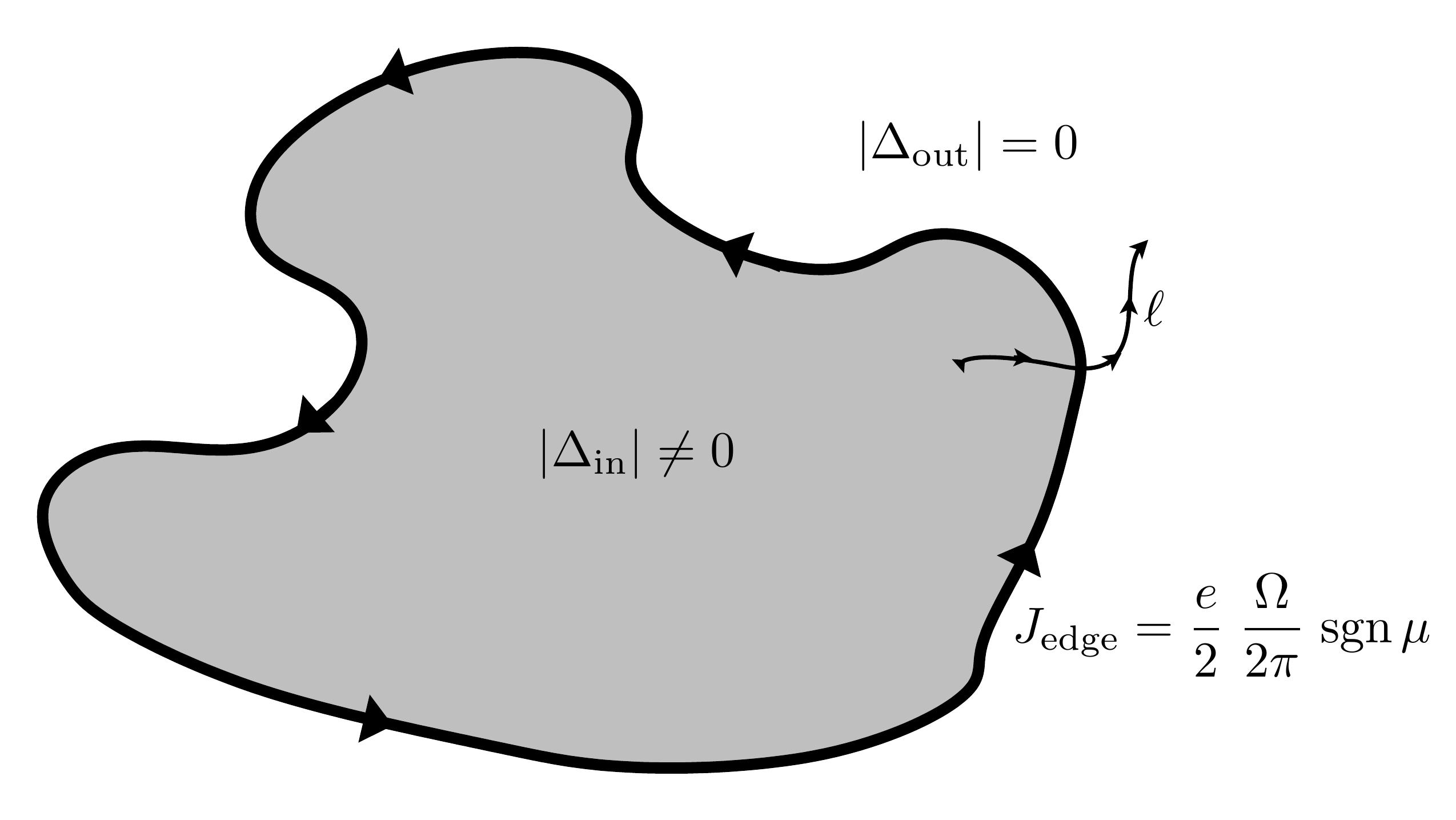}
  \caption{\label{fig:edge_current} 
    Chiral edge current resulting from the out-of-equilibrium
    steady-state arising from the excitation of optical
    phonons at wavevector $\bm K_+$. The direction of the current is
    inverted for $\bm K_-$ phonons, for which $\Omega\to-\Omega$.}
\end{center}
\end{figure}

It follows from \eqref{currentdensity} that the currents flow perpendicular to the gradient of $\abs\Delta$. At the boundary of the sample
$|\Delta|$ must go from constant to zero. Therefore an edge current
should flow parallel to the boundary, within the region where
$\abs\Delta$ varies in space, (see Fig.~\ref{fig:edge_current}). The edge current is given by
\begin{align}\label{edgecurrent}
  J_{\rm edge}
&=\int_{\rm in}^{\rm out}
({\hat{\bm z}}\times d\bm\ell)\cdot \langle\, \bm j\, \rangle\nonumber\\
&=- e\frac{\Omega}{2\pi}
\left(|\chi_{\rm out}|^2-|\chi_{\rm in}|^2 \right)
\;,
\end{align}
where $\bm\ell$ is a path that traverses the boundary. In the interior
of the sample $\abs\Delta$ is non-vanishing, so we can set $\mu\to 0$,
and using Eq.~\eqref{chidefinition} we obtain that $|\chi_{\rm
  in}|^2\to 1/2$. Outside the sample, $\abs\Delta\to 0$ and $\abs\mu\to\infty$.  Depending on whether $\mu>0$ or $\mu<0$ we obtain
$|\chi_{\rm out}|^2\to 0$ or $1$, respectively. Therefore, we arrive
at the edge current
\begin{align}\label{edgecurrent}
  J_{\rm edge}
&=\frac{e}{2}\;\frac{\Omega}{2\pi}\;\text{sgn}\,\mu\;.
\end{align}
The linear relation between $J_{\rm edge}$ and $\Omega$ has a
quantized coefficient. Note that because $\Omega=2\pi/T$, where $T$ is
the rotation period, the current $J_{\rm edge}$ carries a fractional
charge $\pm e/2$ per rotation cycle \footnote{When spin is included,
  \eqref{edgecurrent} acquires a factor of 2, so that a charge $e$ is
  pumped per cycle.}. This chiral current at the boundary of the
steady state bulk insulator is a topological property of the
out-of-equilibrium system; the currents are quantized and protected
against details at the edge, including disorder.

\begin{figure}[htbp]
\begin{center}
  \includegraphics[angle=90,width=.37\textwidth]{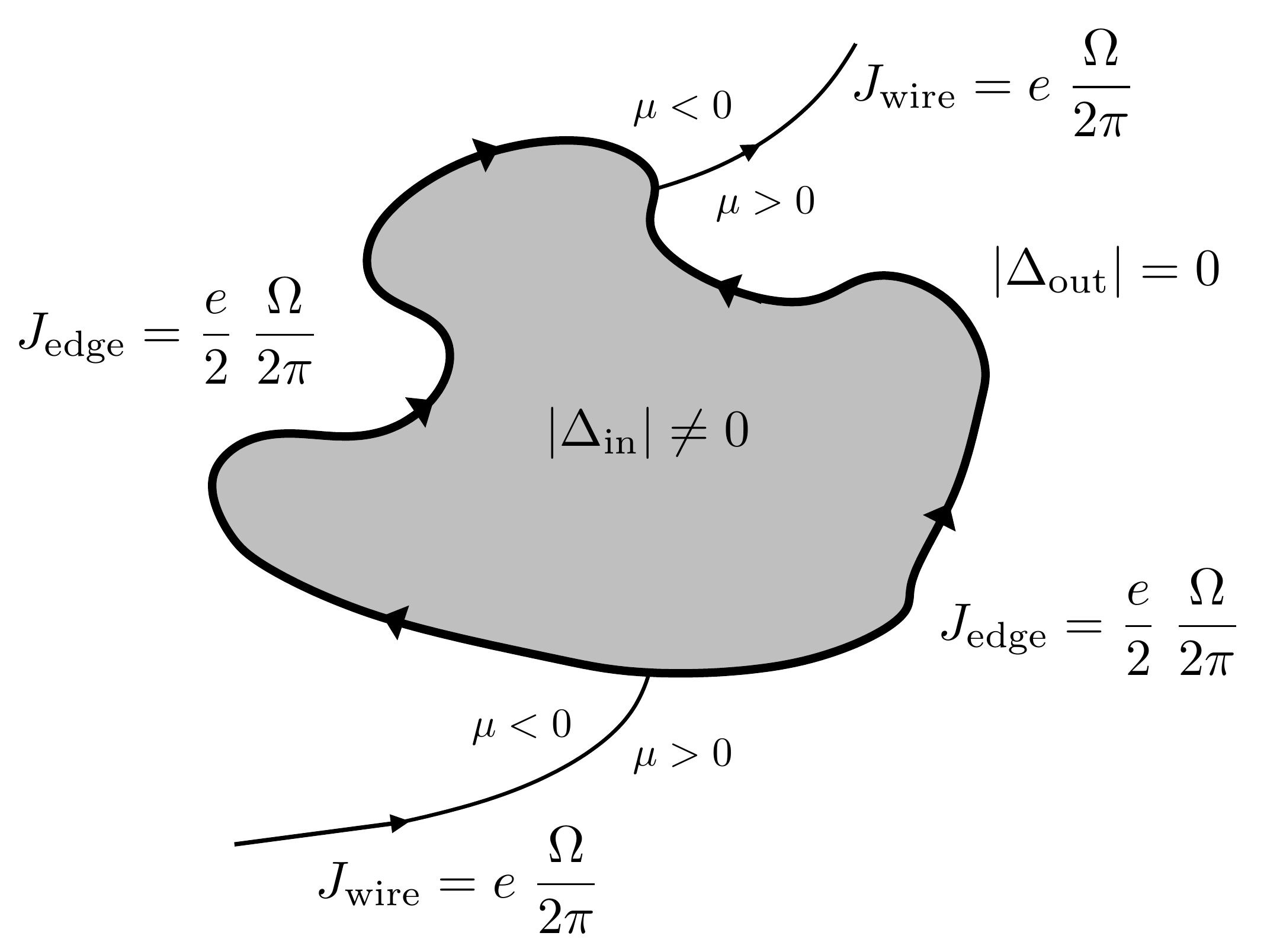}
  \caption{\label{fig:edge_current_wire} Currents in the presence of
    domain walls between regions with $\mu>0$ and
    $\mu<0$. The edge currents have opposite chiralities to either
    side of the wires. The current pumped per cycle is an
    integer multiple of $e$, while a fraction $e/2$ goes around each
    side during the cycle.}
\end{center}
\end{figure}

The chirality of the edge currents depends on whether the $\bm K_+$ or $\bm K_-$ phonon mode is excited.  However, the
chirality of the current also depends on $\text{sgn}\,\mu$.  We now offer a physical explanation of this fact.  The mass $\mu$ was included in the Lagrangian \eqref{currentslagrangian} as a means of terminating the sample with an insulating region. In a physical graphene flake, our findings therefore indicate that the sign of
the edge current is determined by the specific shape of the
sample. Notice that the direction of the current obtained from the
field theory cannot change unless $\mu$ changes sign outside
the sample. But if this is the case, there will be domain walls
separating these regions that support gapless modes. Indeed, these walls serve as quantum wires~\cite{wires} attached to the sample, as shown in
Fig.~\ref{fig:edge_current_wire}. The direction of the edge currents
reverses at the contacts, as shown in the figure. Current conservation requires that currents of magnitude $J_{\rm wire}=
e\,{\Omega}/{2\pi}$ flow in the wires, splitting equally at the
contacts and traveling around the edges of the sample. The graphene flake in this scenario becomes a  pump~\cite{Thouless_pump} that transports a charge $e$ per rotation period $T$.


The next observation concerns zero modes in graphene, which are supported in the presence of vortices in the order parameter
$\Delta$~\cite{changyu2007}. An external chiral gauge potential
$\bm A_5$ was added to render finite the vortex energies, thereby deconfining
them~\cite{chiralgauge}. Such a vortex background can also exist in our time-dependent scenario. In the co-moving
frame this involves adding $A_{5,0} =- \Omega/2$ to the static
problem. We find that zero-energy modes persist both with and without
$\bm A_5$, consistent with the findings of \cite{franz,zeromodes}.

Our final observation concerns the size of the gap that can be
achieved by excitation of the optical phonon modes at $\bm
K_\pm$. From Eq.~\eqref{Delta} we obtain that
$|\Delta|=3\alpha\,t|c_\pm|/d$, where $|c_\pm|/d$ measures
the relative displacement of the atoms from their equilibrium positions
due to the phonons and is controlled by the intensity of the
excitations. Using $\alpha\approx 3.7$ and $t\approx 2.8\, eV$ for
graphene, one obtains for a relative displacement
$|c_\pm|/d\approx 0.04\%$ that $2\abs{\Delta}\approx 0.025$ eV, corresponding to room temperature scales.


In summary, we have illustrated a mechanism for opening
a tunable Kekul\'e gap in graphene by exciting an optical phonon mode at $\bm K_+$ or $\bm K_-$. This gap corresponds to a complex-valued order
parameter $\Delta$ in the continuum theory that rotates in time with frequency $\Omega$. The time dependence of $\Delta$ is completely removable by a gauge
transformation which has no effect on bath degrees of freedom and
leaves the current operators unaffected. The electric response of the
system is therefore equivalent to that of one with a static gap.
Furthermore, the system is found to support chiral quantized currents that
are localized in regions where $\abs\Delta$ varies spatially.  In particular, there are edge currents whose chirality depends on the shape of the sample and on which of the $\bm K_\pm$ phonon modes is excited.


We thank Jerome Dorignac, who participated in an early stage of this
investigation, for useful discussions.  We also acknowledge helpful conversations with Michael El-Batanouny, Bennett Goldberg, Colin Howard, Alex Kitt, Sebastian Remi, and Anna Swan.  This work is supported by
DOE grants DEF-06ER46316 (CC) and -91ER40676 (S-Y~P), by the DARPA-QuEST program (C-Y~H), and by the Deutsche
Forschungsgemeinschaft SPP 1459 and the Alexander von Humboldt Foundation (SVK).  DKC and SVK acknowledge the hospitality of the KITP, through grant NSF PHY11-25915, during its ``Physics of Graphene" program.

\bibliography{rotating_kekule_paper_main}

\end{document}